# A Microcantilever-based Gas Flow Sensor for Flow Rate and Direction Detection

Yu-Hsiang Wang[1], Tzu-Han Hsueh[1], Rong-Hua Ma[2], Chia-Yen Lee[1*],
Lung-Ming Fu[3], Po-Cheng Chou[4], Chien-Hsiung Tsai[5]

[1*]Department of Mechanical and Automation Engineering, Da-Yeh University, Changhua, Taiwan.
[2]Department of Mechanical Engineering, Chinese Military Academy, Kaohsiung, Taiwan.
[3]Department of Materials Engineering, National Pingtung University of Science and Technology, Pingtung, Taiwan.
[4]Department of Interior Design, Shu-Te University of Science and Technology, Kaohsiung, Taiwan
[5]Department of Vehicle Engineering, National Pingtung University of Science and Technology, Pingtung, Taiwan.

*Abstract-* The purpose of this paper is to apply characteristics of residual stress that causes cantilever beams to bend for manufacturing a micro-structured gas flow sensor. This study uses a silicon wafer deposited silicon nitride layers, reassembled the gas flow sensor with four cantilever beams that perpendicular to each other and manufactured piezoresistive structure on each micro-cantilever by MEMS technologies, respectively. When the cantilever beams are formed after etching the silicon wafer, it bends up a little due to the released residual stress induced in the previous fabrication process. As air flows through the sensor upstream and downstream beam deformation was made, thus the airflow direction can be determined through comparing the resistance variation between different cantilever beams. The flow rate can also be measured by calculating the total resistance variations on the four cantilevers.

## I. Introduction

Flow measurement is a necessary task in such diverse fields as medical instrumentation, process control, environmental monitoring, and so forth. Many previous studies have demonstrated the successful applications of MEMS techniques to the fabrication of a variety of flow sensors capable of detecting both the flow rate and the flow direction [1-4]. Broadly speaking, MEMS-based gas flow sensors can be categorized as either thermal or non-thermal, depending upon their mode of operation.

Many thermally-actuated gas flow sensors apply some forms of resistors to evaluate local temperature changes [1-3]. In such sensors, the temperature difference between the sensing resistors varies as the gas flow passes through the sensor, and the comparison of the resistance variations among each resistors enables the gas flow direction to be determined. Adamec *et al.* [1] fabricated a multi-axis hotwire anemometer with four thermoresistor to evaluate flow direction with a power consumption of 25 mW. Kim *et al.* [2] presented a circular-type thermal flow sensor consists of a heater and circular-type temperature sensors to detect flow direction and velocity. However, the power consumption of this device exceeded 80 mW. Recently, non-thermal gas flow meters have been developed. These devices typically have the advantages of lower power consumption and an improved potential for integration with other sensors. Ozaki [4] fabricated an air flow sensor modeled on windreceptor hairs of insects which can detect low velocity of air flow, but the maximum measurable flow rate was just 2 ms$^{-1}$. The practicality of the device for flow rate sensing applications was severely limited. Wang *et al.* [5-7] fabricated an air flow sensor with a free-standing microcantilever structure which has a high sensitivity (0.0284 Ω/ms$^{-1}$) and a high velocity measurement limit (45 ms$^{-1}$). In the current study, four cantilever beams are formed and the measured values of four piezoresistors on them are compared to determined the airflow direction. The flow rate can also be measured by calculating the total resistance variations on the four cantilevers.

## II. Design And Fabrication

### A. Shape Design

As shown in Fig. 1, the electrodes used to connect the piezoresistor to the external LCR meter are fabricated from gold since gold has a lower resistivity than platinum, and hence the resistance effect of the leads is reduced. The resistance R of the piezoresistor is given by

$$R = \frac{l_R \rho}{A} = \frac{l_R \rho}{wt} \quad (1)$$

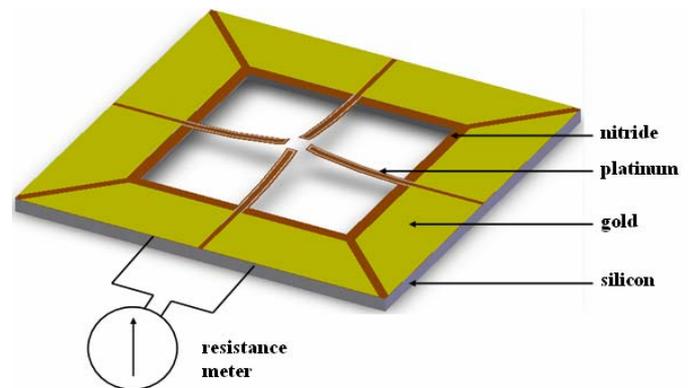

Figure 1 Schematic illustration of gas flow sensor.

where $l_R$ is the resistor length, $ρ$ is the resistivity, $A$ is the cross-sectional area, $w$ is the width and $t$ is thickness. The derivative of (1) is

$$\frac{dR}{R} = \frac{d\rho}{\rho} + \frac{dl_R}{l_R} - \frac{dw}{w} - \frac{dt}{t} \quad (2)$$

where defines $\varepsilon_l = \frac{dl_R}{l_R}$, $dl = \Delta l$, $dw = \Delta w$ and $dt = \Delta t$,

then it could be rearranged to $\frac{dR/R}{\varepsilon_l} \approx (1+2\nu)$ (3)

As shown in Fig. 2, when air flows over the cantilever structure, the beams are deflected that causing a change in the cross-sectional area, and hence the resistance of the platinum resistor changes. The corresponding airflow velocity and direction can then be evaluated and determined simply by measuring the resistance change using an external LCR meter.

*B. Shape Design*

Fig. 3 indicates the configuration and dimensions of the cantilever beams and platinum resistors considered in this study. In Fig. 3(a), the periphery of the black line indicates the cantilever beams. In Fig. 3(b) the black netted area corresponds to the platinum resistors. All units are *μm*. As shown in the figures, the gas flow sensor is assembled with four cantilever beams that perpendicular to each other and manufactured with piezoresistors on each micro-cantilever.

*C. Fabrication*

Fig. 4 presents an overview of the fabrication process used for the current micromachined gas flow sensor. Initially, a low-stress nitride layer of thickness 1.0 μm was deposited on either side of a 500 μm thick double-side-polished silicon wafer. A thin layer of Cr (0.03 μm) was then deposited on the upper surface of the wafer to form an adhesion layer for the subsequent deposition of the platinum resistors. The platinum resistors (0.1μm thickness) were deposited using an electron-beam evaporation process. The same technique was then used to deposit a layer of Au (0.1μm thickness) on the resistors to serve as an electrode and to provide electrical leads. To form the micro-cantilever beams, the cantilever structure and a back-etching nitride mask were patterned in $SF_6$ RIE plasma. In the etching process, the cantilever structure was released in a KOH etchant (40 wt %, 85 °C, supplied by J. T. Baker). Following the etching process, the total micro-cantilever beam thickness was 20 μm (nitride: 1μm / silicon: 19 μm) and the cantilever beams exhibited an upward bending effect as a result of the stress relaxation caused by the thermal stress induced in the heating and

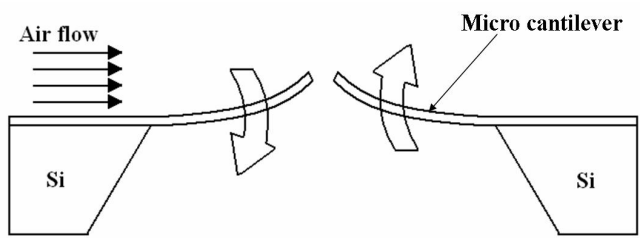

Figure 2 Diagram of gas flow sensor during sensing operation.

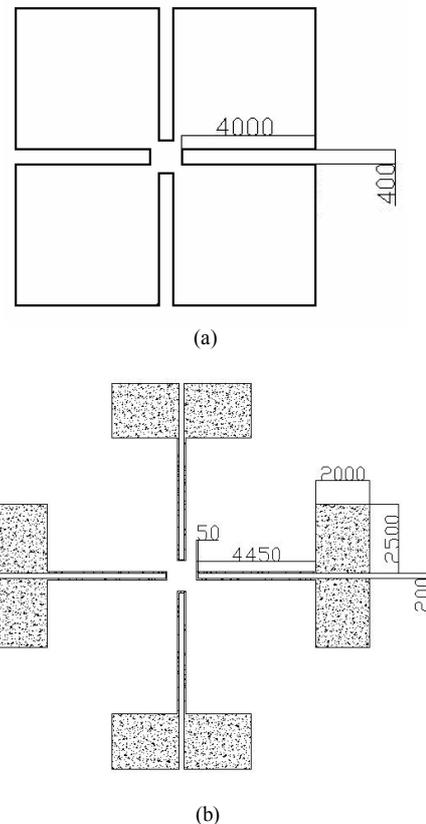

(a)

(b)

Figure 3 Configuration and dimensions of (a) cantilever beams and (b) platinum resistors.

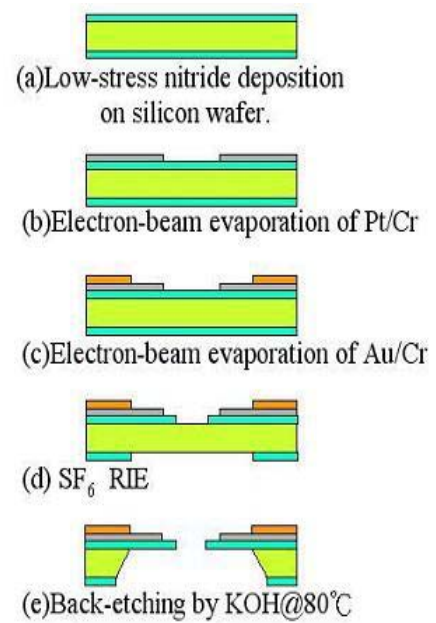

Figure 4 Overview of fabrication process employed for proposed flow sensor.

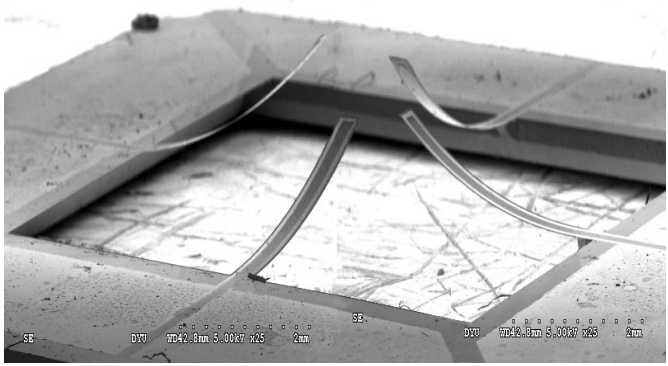

Figure 5 SEM image of gas flow sensor.

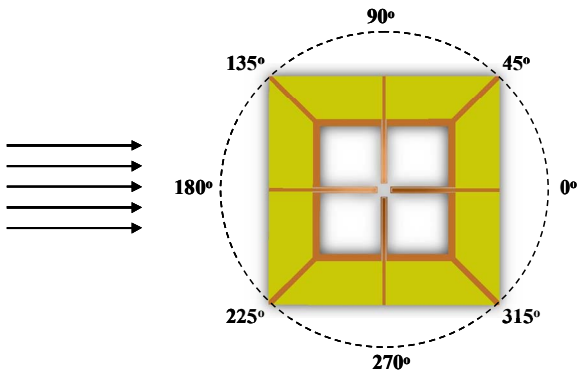

Figure 6 The relative position between the initial air flow direction and the sensor.

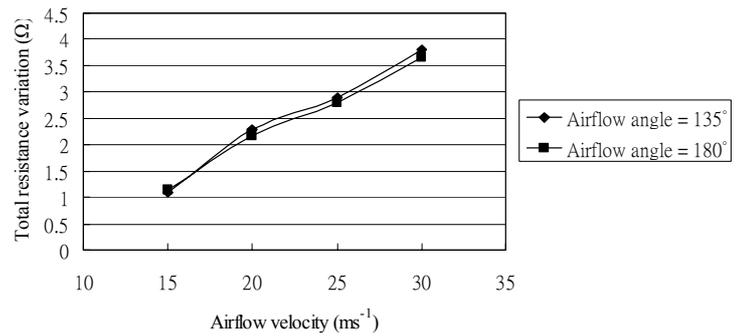

Figure 7 The relationship between the airflow velocity and the total resistance variation when airflow direction was 135 °, and 180 °.

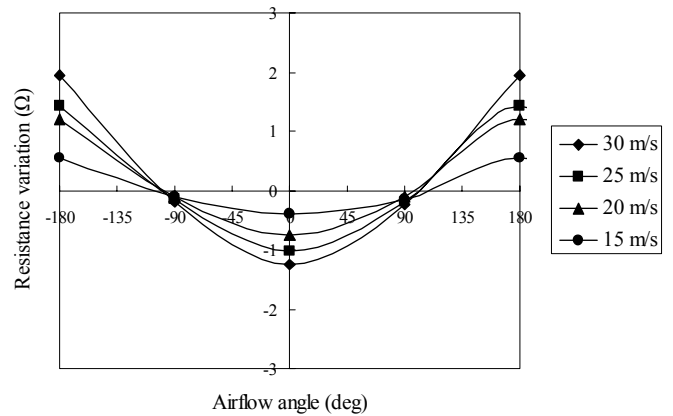

Figure 8 The relationship between the airflow angle and the resistance variation when airflow velocity was (a)15 ms$^{-1}$ [○], (b)20 ms$^{-1}$ [Δ], (c)25 ms$^{-1}$ [□] and (d) 30 ms$^{-1}$ [◊].

cooling steps performed before etching. This phenomenon causes an initial beam deformation position which assists the flow rate sensing function. To increase the reliability of the micro-cantilever, a small thickness of silicon was retained to act as an understructure. Fig. 5 presents a scanning electron microscope (SEM) image of a typical freestanding structure fabricated in this study.

### III. RESULTS AND DISCUSSION

A systematic investigation was conducted into the performance of the fabricated micro gas flow sensors. The characterization of the sensors was carried out with a rotary table (LC-PR60, TanLian E-O Co., Ltd., Taiwan) in a wind tunnel, and the relative position between the air flow direction and the sensor is shown in Fig. 6. The variation in the sensor resistance as the airflow passed over the cantilever was measured using an LCR meter (WK4230, Wayne Kerr Electronics Ltd., Taiwan). For reference purposes, a Pitot tube flow meter was also used to measure the velocity of the airflow.

#### A. Flow Rate

For the of measurement of the flow rate in the wind tunnel, the relationship between the sum of the absolute value of resistance variation in all the four piezoresistors on the beams and flow rate is shown in Fig. 7. It shows that as the airflow direction was at 180 ° or 135 ° direction, the sum of the absolute values of the resistance variation was equal and it just depended on the flow rate.

#### B. Flow Direction

For the measurement of the flow direction, a strong dependence of the measurement was found on the geometrical setup of the sensor axes. It is necessary to assure the cantilevers precise location with the angle 90°. As the airflow pass through the gas flow sensor from 180 °, the largest resistance variation was caused by the downwind cantilever deformation, and the least resistance variation was caused by upwind cantilever deformation. The resistance variation of cantilever beams that perpendicular to the airflow direction are almost equal to zero. The measured values at different flow rate, 30, 25, 20 and 15 m/s, are shown in Fig. 8.

### IV. CONCLUSIONS

The current study develops a MEMS-based gas flow sensor featuring a free-standing micro-cantilever structure. In the sensing operation, the air flow velocity is detected by measuring the change in resistance of the piezoelectric resistors deposited on the cantilever beams as the beam deforms under the effect of the passing airflow and the airflow direction can be obtained by comparing the resistance variation difference between the upstream and downstream cantilever beams to evaluate the airflow angle.

ACKNOWLEDGMENT

The authors would like to thank the financial support provided by the National Science Council in Taiwan (NSC 96-2221-E-212-037 and NSC 95-2218-E-006-022).